\documentclass{article}

\usepackage{arxiv}

\usepackage[utf8]{inputenc} 
\usepackage[T1]{fontenc}    
\usepackage{hyperref}       
\usepackage{url}            
\usepackage{booktabs}       
\usepackage{amsfonts}       
\usepackage{nicefrac}       
\usepackage{microtype}      
\usepackage{lipsum}
\usepackage{graphicx}

\usepackage{natbib}
\usepackage{lastpage}
\usepackage{tikz}
\usepackage[english]{babel}
\usepackage{cleveref}

\usepackage{tkz-graph}
\usepackage{listings}

\definecolor{codegreen}{rgb}{0,0.6,0}
\definecolor{codegray}{rgb}{0.5,0.5,0.5}

\definecolor{backcolour}{RGB}{245,248,250}
\definecolor{emph}{RGB}{166,88,53}
\definecolor{nightblue}{RGB}{9,49,105}
\definecolor{keywords}{RGB}{207,33,46}
\definecolor{lightpurple}{RGB}{130,81,223}

\lstdefinestyle{mystyle}{
    backgroundcolor=\color{backcolour},   
    commentstyle=\color{codegreen},
    keywordstyle=\color{keywords},
    stringstyle=\color{nightblue},
    basicstyle=\fontsize{9}{10}\ttfamily,
    breakatwhitespace=true,         
    breaklines=true,                 
    captionpos=b,                    
    keepspaces=true,                 
    numberstyle=\tiny\color{codegray},
    numbersep=2pt,                  
    showspaces=false,                
    showstringspaces=false,
    showtabs=false,                  
    tabsize=2,
    emph={dspy},
    emphstyle={\color{lightpurple}},
    linewidth=1\columnwidth,
    frame=tb,    
    xrightmargin=0pt,
    xleftmargin=0.23cm,
    numbers=left,
    aboveskip=0.2cm,
    belowskip=0.1cm,
}

\usetikzlibrary{
    arrows,
    arrows.meta,
    fadings,
    patterns,
    positioning,
    shapes.geometric
}

\usetikzlibrary{arrows,topaths,calc}
\usetikzlibrary{positioning}
\usetikzlibrary{shadows}
\usetikzlibrary{shapes}
\usetikzlibrary{backgrounds}
\usetikzlibrary{decorations.pathreplacing}

\graphicspath{ {./images/} }

\title{OWLAPY: A Pythonic Framework for OWL Ontology Engineering}

\author{
 Alkid Baci, \ Luke Friedrichs, \ Caglar Demir, \ Axel-Cyrille Ngonga Ngomo \\
  Department of Computer Science \\
  Paderborn University \\
  Warburger Str. 100, 33098 Paderborn, Germany \\
  \texttt{\{alkid.baci, caglar.demir, axel.ngonga\}@upb.de} \\
  \texttt{lukef@mail.uni-paderborn.de} \\
}

\begin{document}
\maketitle
\begin{abstract}
In this paper, we introduce OWLAPY, a comprehensive Python framework for OWL ontology engineering. 
OWLAPY streamlines the creation, modification, and serialization of OWL 2 ontologies. 
It uniquely integrates native Python-based reasoners with support for external Java reasoners, offering flexibility for users. 
OWLAPY facilitates multiple implementations of core ontology components and provides robust conversion capabilities between OWL class expressions and formats such as Description Logics, Manchester Syntax, and SPARQL. 
OWLAPY allows users to define custom workflows to leverage large language models (LLMs) in ontology generation from natural language text. 
OWLAPY serves as a well tested software framework for users seeking a flexible Python library for advanced ontology engineering, including those transitioning from Java-based environments.
It is publicly available on GitHub \footnote{\url{https://github.com/dice-group/owlapy}} and on Python Package Index (PyPI) \footnote{\url{https://pypi.org/project/owlapy/}}, with over 50,000 downloads at the time of writing.
\end{abstract}

\section{Introduction}
The Web Ontology Language (OWL) plays a key role in modeling and representing complex domains with ontologies \citep{bechhofer2009owl}. It provides the ability to expressively formulate and formalize knowledge structures. Hence, OWL is at the core of technologies such as the Semantic Web~\citep{antoniou2004semantic}, biomedical informatics~\citep{cimino2006practical, hoehndorf2011common, horridge2014webprotege}, and artificial intelligence (AI) driven reason-based systems \citep{trinh2024solving, herron2025potential}.
Traditionally, OWL software development and OWL-based reasoning have been dominated by the Java ecosystem through libraries such as OWLAPI~\citep{horridge2011owl}, which has become the de facto standard for manipulating OWL ontologies. However, as Python continues to gain popularity in various domains including AI, Data Science, and Machine Learning, the need for a native Python interface to OWL becomes increasingly evident \citep{raschka2020machine}. \textbf{OWLAPY} is a Python framework that addresses this gap by providing an interface similar in a fashion akin to the Java-based OWLAPI~\citep{horridge2011owl}. 
It allows users to programmatically create, manipulate, and reason over OWL ontologies entirely within the Python environment. By mirroring key concepts from OWLAPI, OWLAPY offers a familiar and intuitive experience for developers transitioning from Java while making ontology management accessible to Python practitioners who prefer or require a Python-native solution.

\section{OWLAPY}
\label{sec:owlapy}

\begin{figure}[htb]
    \centering
\includegraphics[width=\textwidth]{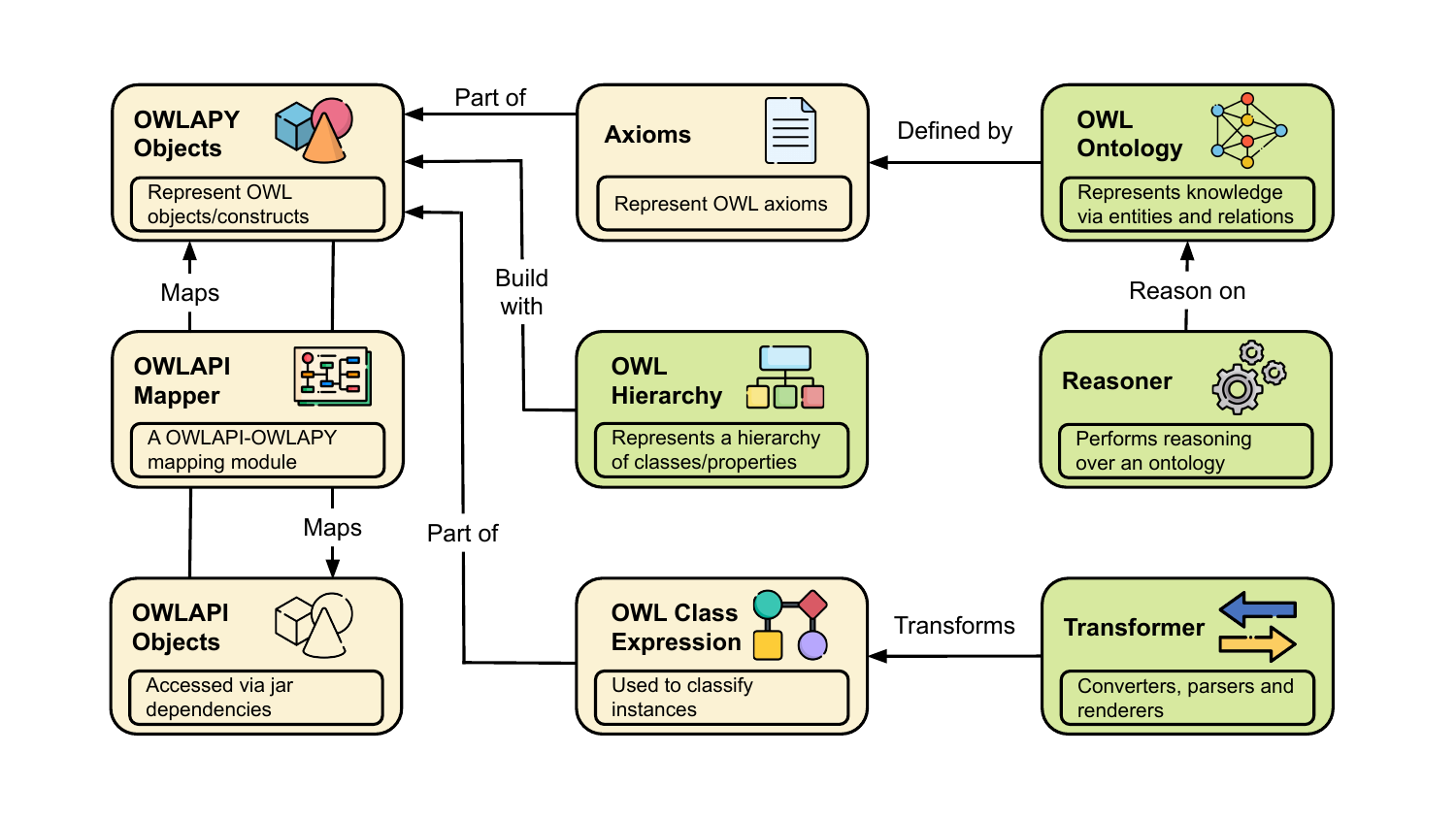}
    \caption{OWLAPY architecture.  Green rectangles show the top-level components, whereas beige rectangles show sub-components. Some components are adapted from \cite{JMLR:v26:24-1113}.}
    \label{OWLAPY_models}
\end{figure}

\Cref{OWLAPY_models} shows the high-level architecture of OWLAPY. The \textbf{Ontology module} in OWLAPY is designed for representing and manipulating OWL ontologies within the Python ecosystem. It allows users to load ontologies from various standard serialization formats into memory. Once loaded, ontologies can be programmatically modified by adding or removing axioms, and subsequently saved back to disk in the desired format. The model offers convenient access to core components of the ontology signature, such as classes, properties, and individuals. Although still recent, OWLAPY also supports the automatic generation of ontologies from natural language text. This functionality is inspired by the approach proposed by \cite{edge2024local} in which entities and their interrelations are extracted using an LLM, and subsequently typed either from a predefined set or through LLM-generated classes. The extracted information is represented in the form of RDF triples, and the system also accommodates triples involving numeric values, thus enabling a degree of semantic richness in the generated ontologies.

Reasoning capabilities are provided through the \textbf{Reasoner module}, which supports native close-world reasoning, embedding-based reasoning and integration with established Java-based OWL reasoners. In particular, OWLAPY introduces the concept of a SyncReasoner, which forwards reasoning tasks to high-performance DL reasoners such as HermiT~\citep{Glimm2014HermiT}, Pellet~\citep{Sirin2007Pellet}, etc., via the JPype~\footnote{\url{https://github.com/jpype-project/jpype}} bridge to the OWLAPI. These reasoners enable open-world, description logic-based inference, including tasks such as subsumption checking, consistency verification, and instance retrieval for complex class expressions.
In addition, OWLAPY provides an embedding-based reasoner (EBR), a native Python model that leverages knowledge graph embeddings for inference, which instead of strict logical deduction, uses the model's scoring function to approximate reasoning tasks, such as retrieving instances of class expressions.

The \textbf{OWL object modeling} approach in OWLAPY adheres closely to the \textit{OWL 2 Structural Specification and Functional-Style Syntax}~\footnote{\url{https://www.w3.org/TR/owl2-syntax/}}, allowing for a faithful representation of OWL constructs. The framework models core OWL entities—such as classes, object properties, data properties, individuals, and literals—as first-class Python objects. These entities form the building blocks for expressing complex class and property expressions, while the axioms module enables users to assert and manipulate logical statements within an ontology

To facilitate readability and interoperability, OWLAPY includes a set of \textbf{conversion utilities} or \textbf{transformers} that translate class expressions between Description Logic (DL) syntax, Manchester syntax, and their programmatic representations. Class expressions can also be converted to SPARQL query which can be further used to query triplestores. OWLAPY has support for SWRL where parsing of string-formatted rules is also possible. Additionally, our framework introduces the \textit{OWLAPIMapper} module, which provides seamless bidirectional mapping between OWLAPI and OWLAPY object structures. This integration leverages the JPype library to enable direct interaction with Java classes, thus one can use OWLAPI directly within our framework.

\begin{figure*}[h]
    \centering
    \small
    \includegraphics[width=0.75\textwidth]{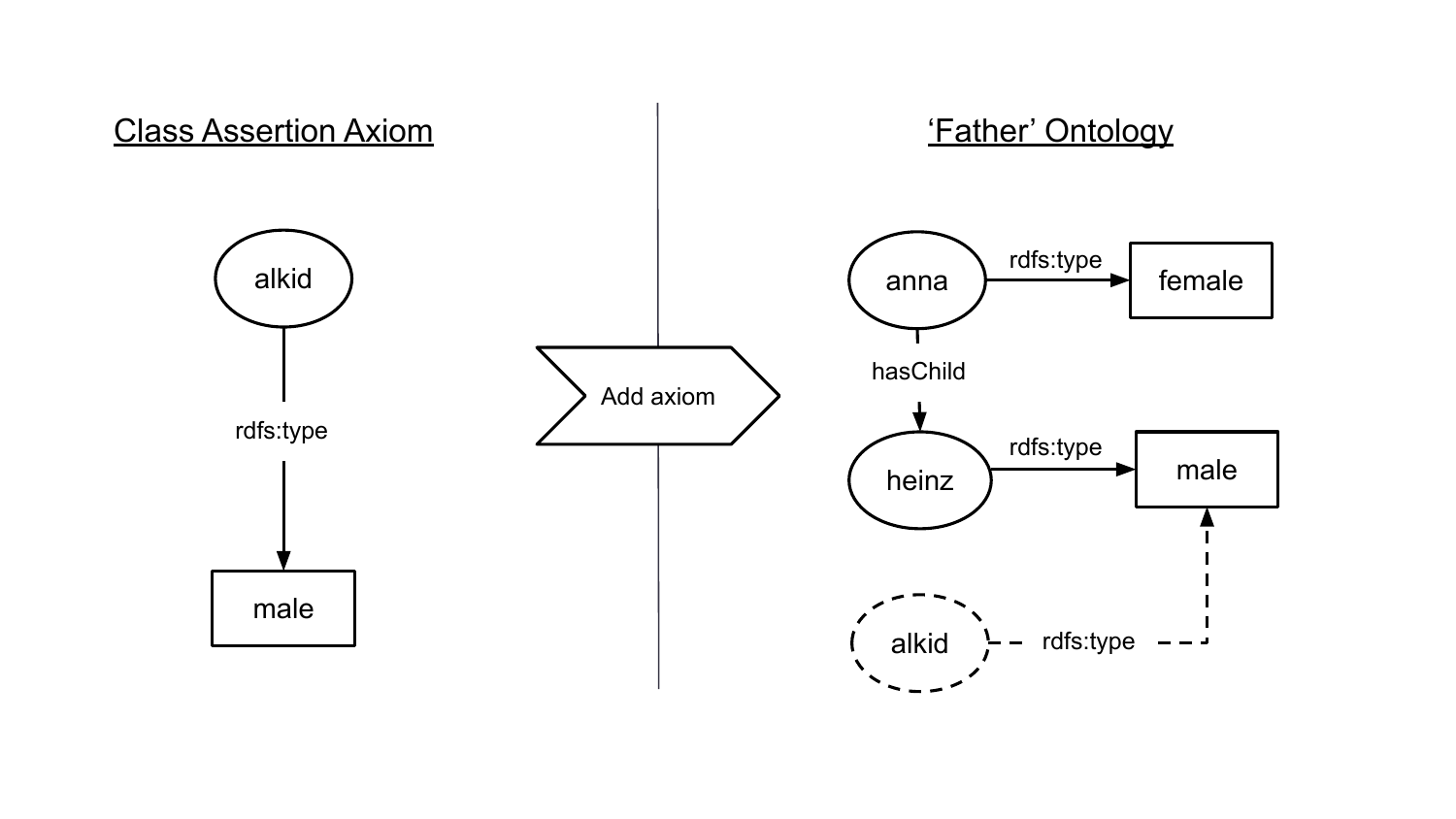}
    \caption{Ontology engineering example as a result of the successful execution of the code in \Cref{sec:owlapy}. Only a portion of the 'father' ontology is displayed. The newly added individual and the class assertion, are shown in dotted lines.}
    \label{fig:ontology_engineering_example}
\end{figure*}

In the code below we show a simple example where a local ontology (the '\textit{father}' ontology) is loaded and a new individual is added. A type is also assigned to this new individual. The updated ontology is then saved for further use. \Cref{fig:ontology_engineering_example} visualizes the successful execution of this code.

\begin{samepage}
\begin{lstlisting}[language=Python,breaklines=true,showstringspaces=false,literate={í}{{\'i}}1]
onto = Ontology("path/to/father_ontology.owl")
male = OWLClass("http://example.com/father#male")
alkid = OWLNamedIndividual("http://example.com/father#alkid")
onto.add_axiom(OWLClassAssertionAxiom(alkid, male))
onto.save(path="updated_father_ontology.owl", rdf_format="rdfxml")
\end{lstlisting}
\end{samepage}

\section{Related Work}
The most closely related Python library to OWLAPY is \textbf{Owlready2}~\citep{lamy2017owlready}, which is also employed within our framework. Owlready2 provides ontology-oriented programming capabilities in Python, with an emphasis on direct access to OWL entities and basic reasoning support through integration with Java-based reasoners such as HermiT and Pellet. However, unlike our library, it does not expose an API structure, which limits its suitability for applications requiring fine-grained, axiomatic manipulation of OWL ontologies and machine learning applications. Another widely used Python library in the semantic web ecosystem is \textbf{RDFlib}~\footnote{\url{https://github.com/RDFLib/rdflib}}, which supports general RDF graph manipulation and SPARQL querying but lacks native support for OWL constructs beyond RDF serialization, and importantly does not provide any OWL Reasoner.\textbf{Ontospy}~\footnote{\url{https://github.com/lambdamusic/Ontospy}} is primarily designed for ontology inspection and visualization, and while useful for exploration, it does not provide programmatic access at the level of expressivity and control required for OWL DL reasoning and editing.

\section{Implementation \& Availability}

The OWLAPY library is distributed as open-source software under the MIT license and is publicly available via both GitHub~\footnote{\url{https://github.com/dice-group/owlapy}} and PyPI~\footnote{\url{https://pypi.org/project/owlapy/}}. Comprehensive documentation~\footnote{\url{https://dice-group.github.io/owlapy/index.html}} is maintained and updated with each release, providing users with detailed guidance on the library’s core functionality and usage patterns. At the time of writing, OWLAPY comprises approximately 15,000 lines of code, includes 11 illustrative example scripts demonstrating key features and counts more than 50,000 downloads. The reliability of the implementation is supported by a suite of 165 unit tests developed using Python’s unittest framework. OWLAPY is designed with extensibility in mind, offering a modular architecture that facilitates the integration of new reasoning components, ontology sources, and features-such as syntax transformers.

\section{Use Cases \& Complementary Libraries}
OWLAPY serves as the foundational library for Ontolearn~\citep{JMLR:v26:24-1113}, a framework for learning OWL class expressions over large knowledge graphs which is applied to industrial projects. OntoSample~\citep{Baci2023Accelerating} offers multiple sampling techniques using OWLAPY to efficiently sample ontologies. Multiple works for the task of class expression learning are based on OWLAPY including here Drill~\citep{demir2023drill}, EvoLearner~\citep{heindorf2022evolearner} and  CLIP~\citep{kouagou2022learning}.

\section*{Acknowledgment}

This work has received funding from the European Union’s Horizon Europe research and innovation program within the project ENEXA under grant agreement No 101070305, the Lamarr Fellow Network program by the Ministry of Culture and Science of North Rhine-Westphalia (MKW NRW) within the project WHALE (LFN 1-04), the Deutsche Forschungsgemeinschaft (DFG, German Research Foundation) within project TRR 318/1 2021 – 438445824, the Ministry of Culture and Science of North Rhine-Westphalia (MKW NRW) within the project SAIL under the grant no NW21-059D and the German Federal Ministry of Research, Technology and Space (BMFTR) within the project KI-Akademie OWL under the grant no 01IS24057B.
OWLAPY is additionally used in teaching at the Master’s and Bachelor’s level at Paderborn University, in courses taught by Prof. Axel Ngonga and TU Dresden, in courses taught by Dr. Patrick Koopmann. We also acknowledge the contributions of the following creators for their icons used in \Cref{OWLAPY_models}: freepik and paul-j, available at \url{https://www.flaticon.com}.

\bibliographystyle{unsrt}  
\bibliography{references}

@inproceedings{Baci2023Accelerating,
  author       = {Alkid Baci and
                  Stefan Heindorf},
  title        = {Accelerating Concept Learning via Sampling},
  booktitle    = {{CIKM}},
  pages        = {3733--3737},
  publisher    = {{ACM}},
  year         = {2023}
}

@article{Glimm2014HermiT,
  author       = {Birte Glimm and
                  Ian Horrocks and
                  Boris Motik and
                  Giorgos Stoilos and
                  Zhe Wang},
  title        = {HermiT: An {OWL} 2 Reasoner},
  journal      = {J. Autom. Reason.},
  volume       = {53},
  number       = {3},
  pages        = {245--269},
  year         = {2014}
}

@inproceedings{Heindorf2022EvoLearner,
  author    = {Stefan Heindorf and
               Lukas Bl{\"{u}}baum and
               Nick D{\"{u}}sterhus and
               Till Werner and
               Varun Nandkumar Golani and
               Caglar Demir and
               Axel{-}Cyrille {Ngonga Ngomo}},
  title     = {EvoLearner: Learning Description Logics with Evolutionary Algorithms},
  booktitle = {WWW},
  year      = {2022}
}

@article{Horridge2011OWL,
  title={The {OWL} {API}: A {Java} {API} for {OWL} ontologies},
  author={Horridge, Matthew and Bechhofer, Sean},
  journal={Semantic web},
  volume={2},
  number={1},
  pages={11--21},
  year={2011},
  publisher={IOS Press}
}

@inproceedings{Kouagou2022Learning,
  author       = {N'Dah Jean Kouagou and
                  Stefan Heindorf and
                  Caglar Demir and
                  Axel{-}Cyrille Ngonga Ngomo},
  title        = {Learning Concept Lengths Accelerates Concept Learning in {ALC}},
  booktitle    = {{ESWC}},
  pages        = {236--252},
  publisher    = {Springer},
  year         = {2022}
}

@article{Sirin2007Pellet,
  author       = {Evren Sirin and
                  Bijan Parsia and
                  Bernardo Cuenca Grau and
                  Aditya Kalyanpur and
                  Yarden Katz},
  title        = {Pellet: {A} practical {OWL-DL} reasoner},
  journal      = {J. Web Semant.},
  volume       = {5},
  number       = {2},
  pages        = {51--53},
  year         = {2007}
}

@article{JMLR:v26:24-1113,
  author  = {Caglar Demir and Alkid Baci and N'Dah Jean Kouagou and Leonie Nora Sieger and Stefan Heindorf and Simon Bin and Lukas Bl{{\"u}}baum and Alexander Bigerl and Axel-Cyrille Ngonga Ngomo},
  title   = {Ontolearn---A Framework for Large-scale OWL Class Expression Learning in Python},
  journal = {Journal of Machine Learning Research},
  year    = {2025},
  volume  = {26},
  number  = {63},
  pages   = {1--6},
  url     = {http://jmlr.org/papers/v26/24-1113.html}
}

@book{antoniou2004semantic,
  title={A semantic web primer},
  author={Antoniou, Grigoris and Van Harmelen, Frank},
  year={2004},
  publisher={MIT press}
}

@article{hoehndorf2011common,
  title={A common layer of interoperability for biomedical ontologies based on OWL EL},
  author={Hoehndorf, Robert and Dumontier, Michel and Oellrich, Anika and Wimalaratne, Sarala and Rebholz-Schuhmann, Dietrich and Schofield, Paul and Gkoutos, Georgios V},
  journal={Bioinformatics},
  volume={27},
  number={7},
  pages={1001--1008},
  year={2011},
  publisher={Oxford University Press}
}

@article{horridge2014webprotege,
  title={WebProt{\'e}g{\'e}: a collaborative Web-based platform for editing biomedical ontologies},
  author={Horridge, Matthew and Tudorache, Tania and Nuylas, Csongor and Vendetti, Jennifer and Noy, Natalya F and Musen, Mark A},
  journal={Bioinformatics},
  volume={30},
  number={16},
  pages={2384--2385},
  year={2014},
  publisher={Oxford University Press}
}

@article{cimino2006practical,
  title={The practical impact of ontologies on biomedical informatics},
  author={Cimino, James J and Zhu, Xinxin},
  journal={Yearbook of medical informatics},
  volume={15},
  number={01},
  pages={124--135},
  year={2006},
  publisher={Georg Thieme Verlag KG}
}

@incollection{bechhofer2009owl,
  title={OWL: Web ontology language},
  author={Bechhofer, Sean},
  booktitle={Encyclopedia of database systems},
  pages={2008--2009},
  year={2009},
  publisher={Springer}
}

@article{trinh2024solving,
  title={Solving olympiad geometry without human demonstrations},
  author={Trinh, Trieu H and Wu, Yuhuai and Le, Quoc V and He, He and Luong, Thang},
  journal={Nature},
  volume={625},
  number={7995},
  pages={476--482},
  year={2024},
  publisher={Nature Publishing Group UK London}
}

@article{herron2025potential,
author = {David Herron and Ernesto Jiménez-Ruiz and Tillman Weyde},
title ={On the Potential of Logic and Reasoning in Neurosymbolic Systems Using OWL-Based Knowledge Graphs},

journal = {Neurosymbolic Artificial Intelligence},
volume = {1},
pages = {29498732251320043},
year = {2025},


}

@article{raschka2020machine,
  title={Machine learning in python: Main developments and technology trends in data science, machine learning, and artificial intelligence},
  author={Raschka, Sebastian and Patterson, Joshua and Nolet, Corey},
  journal={Information},
  volume={11},
  number={4},
  pages={193},
  year={2020},
  publisher={Multidisciplinary Digital Publishing Institute}
}

@inproceedings{demir2023drill,
  author = {Demir, Caglar and Ngonga Ngomo, Axel-Cyrille},
  booktitle = {The 32nd International Joint Conference on Artificial Intelligence, IJCAI 2023},
  title = {Neuro-Symbolic Class Expression Learning},

 year={2023}
}

@article{lamy2017owlready,
  title={Owlready: Ontology-oriented programming in Python with automatic classification and high level constructs for biomedical ontologies},
  author={Lamy, Jean-Baptiste},
  journal={Artificial intelligence in medicine},
  volume={80},
  pages={11--28},
  year={2017},
  publisher={Elsevier}
}

@article{edge2024local,
  title={From local to global: A graph rag approach to query-focused summarization},
  author={Edge, Darren and Trinh, Ha and Cheng, Newman and Bradley, Joshua and Chao, Alex and Mody, Apurva and Truitt, Steven and Metropolitansky, Dasha and Ness, Robert Osazuwa and Larson, Jonathan},
  journal={arXiv preprint arXiv:2404.16130},
  year={2024}
}

\end{document}